\documentclass[proceedings]{JHEP}
\usepackage{amssymb}

\newcommand{\be}{\begin{equation}}
\newcommand{\ee}{\end{equation}}
\newcommand{\ba}{\begin{array}}
\newcommand{\ea}{\end{array}}
\newcommand{\bea}{\begin{eqnarray}}
\newcommand{\eea}{\end{eqnarray}}
\newcommand{\beas}{\begin{eqnarray*}}
\newcommand{\eeas}{\end{eqnarray*}}
\newcommand{\beaa}{\begin{equation}\begin{array}{lcl}}
\newcommand{\eeaa}{\end{array}\end{equation}}
\def\bra{\langle}
\def\ket{\rangle}
\def\vv{\!\!\!\!\!\!\!\!\!\!\!\!}
\def\ww{\!\!\!\!\!\!}
\def\ts{\textstyle}
\def\ds{\displaystyle}
\def\sh#1{\setlength{\parindent}{#1}\indent}

\conference{Non-perturbative Quantum Effects 2000}
\title{Null Vectors in Logarithmic Conformal Field Theory}
\author{Michael Flohr\thanks{Research supported by 
  EU TMR network no.\ FMRX-CT96-0012 and the DFG String network 
  (SPP no.\ 1096), Fl 259/2-1.} \\ 
  Institute for Theoretical Physics, University of Hannover \\
  Appellstra\ss e 2, D-30167 Hannover, Germany \\
  E-mail: \email{flohr@itp.uni-hannover.de}}
\abstract{The representation theory of the Virasoro algebra in the
  case of a logarithmic conformal field theory is considered.
  Here, indecomposable representations have to be taken into
  account, which has many interesting consequences. We
  study the generalization of null vectors towards the case of
  indecomposable representation modules and, in particular,
  how such logarithmic null vectors can be used to derive
  differential equations for correlation functions. We show
  that differential equations for correlation functions with
  logarithmic fields become inhomogeneous.}
\begin{document}
  \noindent {\sc During the last} few years, 
  so-called logarithmic conformal field
  theory (LCFT) established itself as a well-defined knew animal in
  the zoo of conformal field theories in two dimensions \cite{lcft}.
  By now, quite a number of applications have been pursued, and
  sometimes longstanding puzzles in the description of certain
  theoretical models could be resolved, e.g.\ the Haldane-Rezzayi state
  in the fractional quantum Hall effect \cite{fqhe}, multifractality, etc.
  (see \cite{app} for examples).
  
  However, the computation of correlation functions within an LCFT
  still remains difficult, and only in a few cases, four-point functions
  (or even higher-point functions) could be obtained explicitly.
  The main reason for this obstruction is that the representation theory
  of the Virasoro algebra is much more complicated in the LCFT case due
  to the fact that there exist indecomposable but non-irreducible
  representations (Jordan cells). This fact has many wide ranging
  implications. First of all, it is responsible for the appearance of
  logarithmic singularities in correlation functions. Furthermore, it
  makes it necessary to generalize almost every notion of (rational) conformal
  field theory, e.g.\ characters, highest-weight modules, null vectors etc.

  Null vectors are the perhaps most important tool in conformal field
  theory (CFT) to explicitly calculate correlation functions. In certain
  CFTs, namely the so-called minimal models, a subset of highest-weight
  modules possess infinitely many null vectors which, in principle, allow
  to compute arbitrary correlation functions involving fields only out
  of this subset. It is well known that global conformal covariance
  can only fix the two- and three-point functions up to constants. 
  The existence of null vectors makes it possible to
  find differential equations for higher-point correlators, incorporating
  local conformal covariance as well.
  This paper will pursue the question, how this can be
  translated to the logarithmic case.

  For the sake of simplicity, we will concentrate on the case where the
  indecomposable representations are spanned by rank two Jordan cells with
  respect to the Virasoro algebra.
  To each such highest-weight Jordan cell $\{|h;1\ket,|h;0\ket\}$ belong
  two fields, and ordinary primary field $\Phi_h(z)$, and its logarithmic
  partner $\Psi_h(z)$. In particular, one then
  has $L_0|h;1\ket = h|h;1\ket + |h;0\ket$, $L_0|h;0\ket = h|h;0\ket$.
  Furthermore, the main scope will lie on the evaluation of four-point
  functions.
  
\section{{\boldmath $SL(2,\mathbb{C})$} Covariance}

  In ordinary CFT, the four-point function is fixed by global conformal
  covariance up to an arbitrary function $F(x,\bar x)$ of the harmonic
  ratio of the four points, $x=\frac{z_{12}z_{34}}{z_{14}z_{32}}$ with
  $z_{ij}=z_i-z_j$. As usual, we consider only the chiral half of the
  theory, although LCFTs are known not to factorize entirely in chiral
  and anti-chiral halfs.
  
  In LCFT, already the two-point functions behave differently, and the most
  surprising fact is that the propagator of two primary fields vanishes,
  $\bra\Phi_h(z)\Phi_{h'}(w)\ket = 0$. In particular, the norm of the
  vacuum, i.e.\ the expectation value of the identity, is zero. On the other
  hand, all LCFTs possess a logarithmic field $\Phi_0(z)$ of conformal
  weight $h=0$, such that with $|\tilde 0\ket = \Phi_0(0)|0\ket$ the
  scalar product $\bra 0|\tilde 0\ket = 1$. More generally, we have
  \bea 
     \bra\Phi_h(z)\Psi_{h'}(w)\ket &=& \delta_{hh'}\frac{A}{(z-w)^{h+h'}}\,,\\
     \bra\Psi_h(z)\Psi_{h'}(w)\ket &=& \delta_{hh'}\frac{
      B - 2A\log(z-w)}{(z-w)^{h+h'}}\,,\nonumber
  \eea
  with $A,B$ free constants. In an analogous way, the three-point functions
  can be obtained up to constants from the Ward-identities generated by
  the action of $L_{\pm 1}$ and $L_0$. Note that the action of the
  Virasoro modes is non-diagonal in the case of an LCFT,
  \bea\label{eq:vir}
  \ww & {\ds L_n \bra\phi_1(z_1)\ldots\phi_n(z_n)\ket =} & \\
  \ww & {\ds \sum_iz^n\left[z\partial_i + (n+1)(h_i+\delta_{h_i})\right]
          \bra\phi_1(z_1)\ldots\phi_n(z_n)\ket} & \nonumber
  \eea
  where $\phi_i(z_i)$ is either $\Phi_{h_i}(z_i)$ or $\Psi_{h_i}(z_i)$ and
  the off-diagonal action is 
  $\delta_{h_i}\Psi_{h_j}(z) = \delta_{ij}\Phi_{h_j}(z)$ and 
  $\delta_{h_i}\Phi_{h_j}(z) = 0$.
  Therefore, the action of the Virasoro modes yields additional terms
  with the number of logarithmic fields reduced by one. This action reflects
  the transformation behavior of a logarithmic field under conformal
  transformations,
  \be
    \phi_h(z) = \left(\frac{\partial f(z)}{\partial z}\right)^h
    \left(1 + \log(\partial_z f(z))\delta_h\right)\phi_h(f(z))\,.
  \ee

  An immediate consequence of the form of the two-point functions and
  the cluster property of a well-defined quantum field theory is that
  $\bra\Phi_{h_1}(z_1)\ldots\Phi_{h_n}(z_n)\ket = 0$, if all fields
  are primaries. Actually, this is only true if a correlator is considered,
  where all fields belong to Jordan cells. LCFTs do contain other primary
  fields, which themselves are not part of Jordan cells, and whose
  correlators are non-trivial. These are the twist-fields, which sometimes
  are also called pre-logarithmic fields. Such fields belong to the
  fermionic sector of the LCFT, and operator product expansions of two
  twist fields will produce contributions from Jordan cells of primary
  fields and their logarithmic partners. However, since the twist fields
  behave as ordinary primaries with respect to the Virasoro algebra, the
  computation of correlation functions of twist fields only can be
  performed as in the common CFT case. The solutions, however, may exhibit
  logarithmic divergences as well. In this paper, we will compute
  correlators with logarithmic fields, instead.

  Another consequence is that
  \bea
    \lefteqn{\bra\Psi_{h_1}(z_1)\Phi_{h_2}(z_2)\ldots\Phi_{h_n}(z_n)\ket}\\
    &=&\bra\Phi_{h_1}(z_1)\Psi_{h_2}(z_2)\Phi_{h_3}(z_3)\ldots\Phi_{h_n}(z_n)
       \ket\nonumber\\ 
    =\ldots &=& \bra\Phi_{h_1}(z_1)\ldots\Phi_{h_{n-1}}(z_{n-1})\Psi_{h_n}(z_n)
    \ket\nonumber\,.
  \eea
  Thus, if only one logarithmic field is present, it does not matter,
  where it is inserted. Note that the action of the Virasoro algebra does
  not produce additional terms, since correlators without logarithmic fields
  vanish. Therefore, a correlator with precisely one logarithmic field
  can be evaluated as if the theory would be an ordinary CFT.

  It is an easy task to find the general form for four-point functions.
  The final expressions are the more complicated the more logarithmic
  fields are present. One obtains
  \bea
    \bra\Psi_1\Phi_2\Phi_3\Phi_4\ket &=& 
      \prod_{i<j}z_{ij}^{\mu_{ij}}F^{(0)}(x)\,,\\
    \bra\Psi_1\Psi_2\Phi_3\Phi_4\ket &=&
      \prod_{i<j}z_{ij}^{\mu_{ij}}\left[F^{(1)}_{12}(x) - 2F^{(0)}(x)
      \log(z_{12})\right]\,,\nonumber\\ 
    \bra\Psi_1\Psi_2\Psi_3\Phi_4\ket &=&
      \prod_{i<j}z_{ij}^{\mu_{ij}}\left[ F^{(2)}_{123}(x)\right.\nonumber\\ 
     \lefteqn{\vv\vv\vv{}-\sum_{1\leq k<l\leq 3}
       \tilde F^{(1)}_{kl}(x)\log(z_{kl}) + 2F^{(0)}(\log(z_{12})\log(z_{13})
       \nonumber}\\
     \lefteqn{\mbox{}+ \log(z_{12})\log(z_{23}) 
                         + \log(z_{13})\log(z_{23}))\nonumber}\\
     \lefteqn{\vv\vv\vv\left.\mbox{}- F^{(0)}(\log^2(z_{12}) 
                      + \log^2(z_{13}) + \log^2(z_{23}))\right]\nonumber\,,}
  \eea
  where we omit the very lengthy expression for 
  $\bra\Psi_1\Psi_2\Psi_3\Psi_4\ket$. Other choices for the logarithmic fields
  are simply obtained by renaming the indices. The correct combinations
  are
  $\tilde F^{(1)}_{ij}(x) = F^{(1)}_{ik}(x) + F^{(1)}_{jk}(x) 
  - F^{(1)}_{ij}(x)$ with $k$ the remaining index of the third logarithmic
  field. Therefore, the full solution for the four-point function of an
  LCFT involves twelve different functions $F^{(r)}_{i_1\ldots i_{r+1}}(x)$,
  $0\leq r\leq 3$.
  In a similar way, one can make an $SL(2,\mathbb{C})$ covariant ansatz
  for a generic $n$-point function of Jordan cell fields. These results
  generalize the expressions obtained in \cite{kausch} for the $h=0$ 
  Jordan cell of the identity field.

\section{Null vectors in LCFT}

  In an earlier work \cite{null}, all null vectors up to level five were
  explicitly computed, 
  which are built on rank two Jordan cell representations.
  A main feature of these null vectors is that they consist of two
  different descendants, i.e.\
  \be\sh{-0.2cm}
    |\chi^{(n)}_{h,c}\ket = \sum_{|\{m\}|=n}L_{-\{m\}}\left( 
      {\beta\vphantom{'}}^{\{m\}}|h;1\ket +
      {\beta'}^{\{m\}}|h;0\ket\right) 
  \ee
  in an obvious multi-index notation. Within a correlator, such a null vector
  will automatically translate into an inhomogeneous differential equation.
  The homogeneous part is the same as for an ordinary level $n$ null field
  descendant of $\Psi_h$, while the inhomogeneity is given as solution of 
  another differential equation, corresponding to a non-trivial descendant of
  $\Phi_h$. On the other hand, if we consider the differential equation
  for a null field on the primary $\Phi_h$, we still end up with an
  inhomogeneous differential equation due to the other logarithmic 
  fields (there must be at least one!) in the correlator.

  Thus, the coefficients ${\beta\vphantom{'}}^{\{m\}}$ are determined as
  functions in $h,c$ by the linear system of equations
  \be
    L_{\{p\}}\sum_{|\{m\}|=n}{\beta\vphantom{'}}^{\{m\}}L_{-\{m\}}|h;1\ket = 0
    \ \ \forall\ \ |\{p\}|=n
  \ee
  in the usual way. Using the commutation relations of the Virasoro algebra,
  these equations are reduced to equations involving solely $L_0$ and the
  central charge, i.e.
  \be
    \sum_{|\{m\}|=n} {\beta\vphantom{'}}^{\{m\}}f_{\{p\},\{m\}}(L_0,C)
    |h;1\ket = 0\,.
  \ee
  Now, due to the off-diagonal part of the
  action (\ref{eq:vir}) of the Virasoro algebra, one gets additional
  contributions proportional to $|h;0\ket$ which have to be canceled by the
  new coefficients ${\beta'}^{\{m\}}$. With
  $L_0|h;1\ket = (h+\delta_h)|h;1\ket$, one can show that these equations take
  the form
  \bea
    \sum_{|\{m\}|=n} {\beta\vphantom{'}}^{\{m\}}f_{\{p\},\{m\}}(h,c)
    |h;1\ket = 0\,,\ \ \ \ \ \ \ \ & & \\
    \sum_{|\{m\}|=n} \!\!({\beta'}^{\{m\}} +
    {\beta\vphantom{'}}^{\{m\}}\partial_h)f_{\{p\},\{m\}}(h,c)
    |h;0\ket &=& 0\,.\nonumber   
  \eea
  A solution to these equations is given by putting
  ${\beta'}^{\{m\}}(h,c=c'(h))
  = \partial_h {\beta\vphantom{'}}^{\{m\}}(h,c=c(h))$ where the condition
  $c(h)=c'(h)$ fixes the possible values of the central charge to a 
  discrete set. 
  Often, the simpler null state conditions $L_p|\chi^{(n)}_{h,c}\ket = 0$ 
  for $p=1,\ldots n$ are used. Although they are
  equivalent to the above conditions in ordinary CFT, they only
  provide sufficient but not necessary conditions in the LCFT case, as
  can already seen at level three.
  
  For example, the conditions for logarithmic null states at level two 
  are firstly the well-known ones for an ordinary level two null state,
  ${\beta\vphantom{'}}^{\{2\}}=-\frac23(2h+1){\beta\vphantom{'}}^{\{1,1\}}$,
  ${\beta\vphantom{'}}^{\{1,1\}}={\it const}$, 
  and $c=2h(5-8h)/(2h+1)$. In addition, the off-diagonal contributions yield
  ${\beta'}^{\{2\}}=-\frac43h{\beta\vphantom{'}}^{\{1,1\}}$, 
  ${\beta'}^{\{1,1\}}=0$, and $c=5-16h$. The two different conditions for
  the central charge have two common solutions, namely
  $(h=-\frac54,c=25)$ and $(h=-\frac14,c=1)$.
  
  \section{Correlation Functions}

  With the generalization of null vectors to the logarithmic case at hand,
  the next question is how to effectively compute correlation functions
  involving fields from non-trivial Jordan cells.
  As an example, we consider a four-point function with such a primary field 
  which is degenerate at level two. To simplify the formul\ae, we fix the 
  remaining three points in the standard way, i.e.\ we consider
  $G_4=\bra\phi_1(\infty)\phi_2(1)\Phi_{h_3}(z)\phi_4(0)\ket$. According to
  (\ref{eq:vir}), the level two descendant yields
  \be\label{eq:null}
    \left[\frac{3\,\partial_z^2}{2(2h_3+1)} + \!\sum_{w\neq z}
    \!\left(
    \frac{\partial_w}{w-z}-\frac{h_w+\delta_{h_w}}{(w-z)^2}\right)\right]G_4=0
    \,.
  \ee
  If there is only one logarithmic field, $\delta_h$ will produce a four-point
  function without logarithmic fields, i.e.\ won't yield an additional term.
  Hence, after rewriting this equation as an ordinary differential equation
  solely in $z$, we can express the conformal blocks in terms of
  hypergeometric functions Putting without loss of generality the
  logarithmic field at infinity, we can rewrite
  \bea
    G_4 &=& z^{p+\mu_{34}}(1-z)^{q+\mu_{23}}F^{(0)}(z)\,,\\
    p   &=& {\textstyle
            \frac16-\frac23h_3-\mu_{34}-\frac16\sqrt{r_4}}\,,\nonumber\\
    q   &=& {\textstyle
            \frac16-\frac23h_3-\mu_{23}-\frac16\sqrt{r_2}}\,,\nonumber\\
    r_i &=& 1-8h_3+16h_3^2+48h_ih_3+24h_i\,,\nonumber
  \eea
  with $F^{(0)}$ being a solution of the hypergeometric system 
  ${}_2F_1(a,b;c;z)$ given by
  \bea
    a &=& {\textstyle
          \frac12-\frac16\sqrt{r_2}-\frac16\sqrt{r_4}-\frac16\sqrt{r_1}
          }\,,\nonumber\\
    b &=& {\textstyle
          \frac12-\frac16\sqrt{r_2}-\frac16\sqrt{r_4}+\frac16\sqrt{r_1}
          }\,,\nonumber\\
    c &=& {\textstyle
          1 - \frac13\sqrt{r_4}
          }\,.
  \eea
  The next complicated case is the presence of two logarithmic fields.
  The ansatz now reads
  \be\sh{-1cm}
     G_4 = z^{p+\mu_{34}}(1-z)^{q+\mu_{23}}\left(
     F^{(1)}_{ij}(z) - 2\log(w_{ij})F^{(0)}(z)\right)\,.
  \ee
  Surprisingly, if the two logarithmic fields are put at $w_2=1$ and $w_4=0$,
  the additional term in the new ansatz vanishes. However, the $\delta_h$
  operators in (\ref{eq:null}) create two terms such that the standard
  hypergeometric equation becomes inhomogeneous,
  \bea
    \!\!\! & {\ds\left[z(1-z)\partial_z^2 + (c-(1+a+b)z)\partial_z - ab\right]
          F^{(1)}_{24}(z)} & \nonumber\\
    \!\!\! & {\ds = \frac{{\textstyle\frac23}(2h_3+1)}{z(1-z)}F^{(0)}(z)\,.} &
  \eea
  The solution of this inhomogeneous equation cannot be given in
  closed form, it involves integrals of products of hypergeometric functions.
  But for special choices of the conformal weights, simple
  solutions can be obtained. The best known LCFT certainly is the
  CFT with central charge $c=c_{2,1}=-2$. The field of conformal weight
  $h=h_{2,1}=1$ in the Kac table possesses a logarithmic partner.
  Choosing all weights in the four-point function to be equal to $h$, we find
  with ${}_2F_1(-4,-1;-2;z) = A(2z-1) + Bz^3(z-2)\equiv Af_1+Bf_2$ the
  solutions
  \bea\label{eq:2sol}
    \lefteqn{F^{(0)}(z) = [z(1-z)]^{-4/3}(Af_1+Bf_2)\,,}\\
    \lefteqn{F^{(1)}_{24}(z) = [z(1-z)]^{-4/3}\left[Cf_1 + Df_2\right.
    \nonumber}\\
    &+& {\textstyle(\frac23(B-2A)f_2-\frac23Af_1)}\log(z)\nonumber\\
    &-& {\textstyle(\frac23(B-2A)f_2-\frac23Af_1)}\log(1-z)\nonumber\\
    &+& {\textstyle\frac19(6z^2-6z-7)Af_1}
    + \left.{\textstyle(-\frac23z^3+\frac{5}{9}f_1)B}\right]\,.\nonumber
  \eea
  Note that $F^{(0)}$ does not depend on which field is the logarithmic
  one (hence the omitted lower index), since only the contraction
  of {\em two\/} logarithmic fields causes logarithmic divergences.
  A nice example for this is the twist field $\mu(z)$ in the $c=-2$ LCFT, 
  which has $h=-1/8$. Although its OPE with itself yields a logarithmic term,
  $\mu(z)\mu(w)\sim \tilde{\mathbb{I}}(w) + \log(z-w)\mathbb{I}$, no
  logarithm shows up in its two-point function. At least four twist fields
  are necessary to get a logarithm in a correlation function, which is
  equivalent to two logarithmic fields, since
  $\tilde{\mathbb{I}}(z)\tilde{\mathbb{I}}(w)\sim
  -2\log(z-w)\tilde{\mathbb{I}}(w) -\log^2(z-w)\mathbb{I}(z)$.

  So far, we have considered correlation functions with logarithmic fields,
  but where the null field condition was exploited for a primary field.
  As mentioned at the beginning of this section, a null vector descendant
  on the full Jordan cell (not on its irreducible subrepresentation) is
  more complicated. For example, the logarithmic partner of the $h=1$ field
  in the $c=-2$ LCFT turns out to be the $h=h_{1,5}$ field in the Kac table.
  In deed, as shown in \cite{null}, there exists a null vector of the form
  \bea\label{eq:l5}
    \lefteqn{|\chi^{(5)}_{h=1,c=-2}\ket} \\
    &=& [{\textstyle\frac{16}{3}}L_{-1}L_{-2}^2 + \textstyle{\frac{52}{3}}
        L_{-2}L_{-3} - 12L_{-1}L_{-4}\nonumber\\
    & & {}+ \textstyle{\frac{148}{3}}L_{-5}]|h;0\ket\nonumber\\
    &+& [L_{-1}^5 - 10L_{-1}^3L_{-2} + 36L_{-1}^2L_{-3} - L_{-1}L_{-4} 
        \nonumber\\
    & & {}+ 16L_{-1}L_{-2}^2 - 40L_{-2}L_{-3} + 160L_{-5}]
        |h;1\ket\nonumber
  \eea
  The first descendant is precisely the same as for a primary field degenerate
  of level five. However, a remarkable fact in LCFT is that the null
  descendant factorizes,
  \bea
    \lefteqn{|\chi^{(5)}_{h=1,c=-2}\ket = (\ldots)|h;0\ket + \mbox{}}\\
    & (L_{-1}^3 - 8L_{-1}L_{-2} + 20L_{-3})(L_{-1}^2 - 2L_{-2})
      |h;1\ket & \nonumber
  \eea
  namely into the level two null descendant times a level three descendant
  which turns out to be the null descendant of a primary field of conformal
  weight $h_{3,1}=3$. Hence, the level two descendant of the logarithmic 
  field is a null descendant only up to a primary field of weight
  $h_{3,1}=h_{1,5}+2$.

  Another important point is that the additional descendant on the primary
  partner is not unique. The typical LCFT case is that the
  logarithmic partner constituting a Jordan cell representation is 
  degenerate of level $n+k$ with $n$ the level where the primary has its
  null state, and $k>0$. Then, the descendant of the primary field is
  determined only up to an arbitrary contribution $\sum_{|\{m\}|=k}
  \alpha^{\{m\}}L_{-\{m\}}|\chi^{(n)}_{h,c}\ket$.
  
  That the $(1,5)$ entry of the Kac table does indeed refer to the
  logarithmic partner of the $h=1$ primary $(2,1)$ field can be seen from
  the solutions of the homogeneous differential equation resulting from
  (\ref{eq:l5}) when there are no off-diagonal contributions. Of course,
  the resulting ordinary differential equation of degree five has, among
  others, the same solutions as the hypergeometric equation above for the
  $(2,1)$ field. These are the correct solutions, if there is no other
  logarithmic field. The other three solutions turn out to have logarithmic
  divergences. Therefore, they cannot be valid solutions for this case,
  but must constitute solutions for a correlator with two logarithmic fields.
  However, in this case one has to take into account that the full null
  state has an additional contribution from the primary partner of the
  $(1,5)$ field. The full inhomogeneous equation reads (with the ``simplest''
  choice for the primary part of the descendant)
  \bea
    0 &=&\left[z^3(1-z)^3\partial^5 + 8z(z-1)(z^2-z+1)\partial^3\right.
          \nonumber\\
    &-& 4(2z-1)(5z^2-5z+2)\partial^2 + 24(2z-1)^2\partial\nonumber\\ 
    &-& \left.48(2z-1)^{\vphantom{2}}\right]F^{(1)}_{34}(z) \nonumber\\
    &+&\left[-{\ts\frac{16}{3}}z(z-1)(2z-1)^2\partial^3\right.\\
    &+&{\ts\frac{44}{3}}(2z-1)(5z^2-5z+2)\partial^2\nonumber\\
    &-&{\ts\frac{8}{z(z-1)}}(57z^4-114z^3+90z^2-333z+5)\partial\nonumber\\
    &+&{\ts\frac{16}{z(z-1)}}\left.(2z-1)(18z^2-18z+5)\right]F^{(0)}(z)
    \nonumber 
  \eea
  in the case of one further logarithmic field put at zero. Similar equations
  can be written down for all three choices $F^{(1)}_{3j}(z)$ as well as
  for higher numbers of logarithmic fields. In general, there is one part
  of the differential equation for $F^{(r)}_{I}$ with $I=\{3,i_1,\ldots,i_r\}$, 
  and the inhomogeneity is given by $F^{(r-1)}_{I-\{3\}}$. It is clear from 
  this that the full set of solutions can be obtained in a hierarchical
  scheme, where one fist solves the homogeneous equations and increases
  the number of logarithmic fields one by one. 

  In the example above, $F^{(0)}$ is given as in (\ref{eq:2sol}). Then the
  inhomogeneity reads $80(3z^2-3z+1)A+16z(z^2-9z+3)B$. With this,
  the solution is finally obtained to be
  \bea
    \lefteqn{F^{(1)}_{34} =  C_1f_1 + C_2f_2
     +  C_3[3f_1\log({\ts\frac{z}{z-1}}) - 6]\nonumber}\\
    &+& C_4[3f_2\log(z-1)-12z^3]\nonumber\\
    &+& C_5[3(f_1+f_2)\log(z) + 12z(z^2-3z+1)]\nonumber\\
    &+& \left[{\ts\frac29}(3f_1-2f_2)\log(z)\right.
                + {\ts\frac29}(7f_1+2f_2)\log(z-1)\nonumber\\
    & & \mbox{} + {\ts\frac{1}{27}}\left.(12z^3-18z^2+32z-1)\right]A\nonumber\\
    &+& \left[{\ts\frac29}(f_2-f_1)\log(z)\right.
                - {\ts\frac29}(4f_1+f_2)\log(z-1)\nonumber\\
    & & \mbox{} + {\ts\frac{1}{27}}\left.(36z^2-6z^3-17f_1)\right]B\,.
  \eea
  As is obvious from the above expression, correlation functions involving
  more than one logarithmic field become quite complicated. Although 
  the two logarithmic fields were chosen to be located at $z,0$, the above
  solution also contains terms in $\log(z-1)$. This is a consequence of
  the associativity of the OPE and duality of the four-point function.

  In principle, the full set of four-point functions can be evaluated in
  this way. Care must be taken with the solutions of the homogeneous
  equation. As indicated above, not all of them might be valid solutions.
  If the correlator does contain only one logarithmic field, then
  there cannot be any logarithmic divergences in the solution. However,
  it is instructive to find the reason, why already the homogeneous
  equation admits logarithmic solutions. Firstly, one should remember that
  a similar situation arises in minimal models. All primary fields come
  in pairs in the Kac table, which are usually identified with each other,
  $(r,s) \equiv (q-r,p-s)$ if the central charge is $c=c_{p,q}$. So, in 
  principle, one and the same correlator can be evaluated by exploiting
  two different null state conditions, which in general will be of 
  different degrees, $rs\neq rs+qp -(qs+pr)$. Therefore, the physical 
  solutions are given by the intersection of the two sets of solutions.

  In the logarithmic case, the typical BPZ argument that only the common
  set of fusion rules can be non-vanishing \cite{BPZ}, has to be modified.
  The $(2,1)$ field has the formal BPZ fusion rules
  $[(2,1)]\times[(2,1)] = [(1,1)] + [(3,1)]$, but the last term must vanish
  due to dimensional reasons, since $h_{3,1} =3 > 2h_{2,1} = 2\cdot 1$.
  On the other hand, one has in a formal way 
  $[(2,1)]\times[(1,5)] = [(1,1)]$, meaning that the OPE of the logarithmic
  field with its own primary partner won't yield a logarithmic dependency.
  Note that a logarithmic field can be considered as the normal ordered
  product of its primary partner with the logarithmic partner of the 
  identity, i.e.\ $\Psi_h(z) = \mbox{:$\Phi_h\tilde{\mathbb{I}}$:}(z)$.
  As long as an OPE of such a field with a primary field is considered,
  one can evaluate it in the usual way, and then take the normal ordered
  product of the right hand side with $\tilde{\mathbb{I}}$, since the
  latter field behaves almost as the identity field with respect to fusion 
  with primary fields.
  But as soon as the OPE of two logarithmic fields is taken, one gets
  a new term: $[(1,5)]\times[(1,5)] = [(1,1)] + [(1,3)] + [(1,5)]$,
  where all terms are omitted which must vanish due to dimensional reasons.
  Now, the (1,3) field $\tilde{\mathbb{I}}$ itself appears in the OPE, which is
  correct because the OPE of two such normal ordered products will 
  involve the well-known OPE $\tilde{\mathbb{I}}(z)\tilde{\mathbb{I}}(w)
  \sim -2\log(z-w)\tilde{\mathbb{I}}(w)$.
  This proves that the logarithmic solutions of the conformal blocks of
  the four-point function can only be valid when sufficiently many
  logarithmic fields are involved.

  This leads back to the above mentioned observation that the null state
  of a logarithmic field of level $n+k$ factorizes into the level $n$ 
  null descendant of its primary partner times the level $k$ null state of
  a primary field of conformal weight $h+n$. Indeed, it is a nice exercise
  to show that the Virasoro modes of the level two null
  descendant, acting on the logarithmic $\Psi_{h=1}$ field, produce 
  a field which transforms as a primary field of conformal weight $h=3$.
  The reason is that the derivative, acting on a logarithmic field, eats
  up the fermionic zero modes. Namely, in
  \bea
  [L_{-n},\Psi_h(z)] = z^n((n+1)h+z\partial)\Psi_h(z)\phantom{mmml} & &\\
  = (n+1)h\Psi_h(z) + \mbox{:$(\partial\Phi_h)\tilde{\mathbb{I}}$:}(z)
  + \mbox{:$\Phi_h(\partial\tilde{\mathbb{I}})$:}(z)\,. & & \nonumber
  \eea
  where the $\delta_h$ part is omitted, the derivative first acts as
  derivative on the primary part of the logarithmic field, and then acts
  on the field $\tilde{\mathbb{I}}$. In the $c=-2$ LCFT this basic logarithmic
  field can be constructed out of two anticommuting scalar fields,
  \be
  \theta^{\alpha}(z) = \sum_{n\neq 0}\theta_n^{\alpha}z^{-n} 
                     + \theta_0^{\alpha}\log(z) + \xi^{\alpha}\,,
  \ee
  $\alpha=\pm$, whose zero modes are responsible for all the logarithms.
  Then $\tilde{\mathbb{I}}(z) = -\frac12\epsilon_{\alpha\beta}
  \mbox{:$\theta^{\alpha}\theta^{\beta}$:}(z)$. Therefore, the derivative
  will eat up zero modes, e.g.\ $\tilde{\mathbb{I}}(0)|0\ket = 
  \xi^+\xi^-|0\ket$ and $\partial\tilde{\mathbb{I}}(0)|0\ket =
  (\theta_{-1}^+\xi^- + \theta_{-1}^-\xi^+)|0\ket$. By considering states,
  one can show that the level two null descendant applied to the state 
  $\Psi_{h=1}(0)|0\ket$ yields a state proportional to a 
  highest-weight state of weight $h=3$.
  
\section{Conclusion}

  Taking into account the proper action of the Virasoro algebra on
  logarithmic fields, i.e.\ working with Jordan cell representations as
  generalizations of irreducible highest-weight representations, allows
  to evaluate correlations functions in LCFT in a similar fashion as in
  ordinary CFT. The main difference is that each $n$-point function represents
  a full hierarchy of conformal blocks involving $r+1=1,\ldots,n$ logarithmic
  fields. The solution of this hierarchy can be obtained step by step,
  where the case with one logarithmic field only is worked out in the same
  way as in ordinary CFT. In each further step, the differential equations
  are inhomogeneous, with the inhomogenity determined by the conformal 
  blocks of correlators with fewer logarithmic fields. A more detailed
  exposition including twist fields will appear elsewhere \cite{soon}. 
  This fills one of 
  the few remaining gaps to put LCFT on equal footing with better known 
  ordinary CFTs such as minimal models.

\acknowledgments
  Many thanks to Philippe Ruelle for getting me interested in the
  mess with inhomogeneous differential equations in LCFTs and for 
  many email discussions and collaboration. 
  During preparation of the final manu\-script,
  we received a draft of \cite{tehran} which partially overlaps and
  partially disagrees with the results reported here.

\end{document}